\documentclass[conference]{IEEEtran}

\pdfoutput=1
\input{macros}

\newcommand{\fullp}[1]{}

\newcommand{\bpr}{\begin{IEEEproof}}
\newcommand{\epr}{\end{IEEEproof}}

\begin{document}

\title{Specification Construction Using Behaviors, Equivalences, and SMT Solvers}

\author{
\IEEEauthorblockN{Paul C. Attie, Fadi Zaraket, Mohamad Noureddine, and Farah El-Hariri}
\IEEEauthorblockA{American University of Beirut, Beirut, Lebanon} 
}

\maketitle

\begin{abstract}
  We propose a method to write and check a specification
  including {\em quantifiers} using behaviors, i.e., 
  input-output pairs.  Our method requires the
  following input from the user: (1) answers to a finite number of
  queries, each of which presents some behavior to the user, who
  responds informing whether the behavior is ``correct'' or not; and
  (2) an ``equivalence'' theory (set of formulae), which represents the users opinion
  about which pairs of behaviors are equivalent with respect to the
  specification; and (3) a ``vocabulary,'' i.e., a set of formulae
  which provide the basic building blocks for the specification to be written.
  Alternatively, the user can specify a type theory and a simple relational 
  grammar,
  and our method can generate the vocabulary and equivalence theories. 
  Our method automatically generates behaviors using a {\em satisfiability
  modulo theory} solver. 


  Since writing a specification consists of formalizing ideas that are
  initially informal, \textit{there must, by definition, be at least
    one ``initial'' step where an informal notion is formalized by the
    user in an ad hoc manner.}  This step is the provision of the
  equivalence theory and vocabulary; we call it the 
  \emph{primitive formalization step}. 

  We contend that it is considerably easier to write an equivalence
  theory and vocabulary than to write a full-blown formal
  specification from scratch, and we provide experimental evidence for
  this claim. We also show how vocabularies can be constructed
  hierarchically, with a specification at one level providing
  vocabulary material for the next level.

\end{abstract}



\remove{
Our method is based on (1) a (given) type theory for the variables
referenced in the specification, and (2) an equiavlence theory defined by the developer, which 
represents the developers opinion about which pairs of behaviors are equivalent w.r.t. the specification, 
and (3) a vocabulary of lower level concepts. This vocabulary must be ``adequate'' in that
it enables good and bad behaviors to be distinguished. We present two algorithms: one to construct a specification from
an adequate vocabulary, and one to check if a vocabulary
is adequate, and to correct it if it is not.
}

\section{Introduction}
\label{s:intro}

The derivation of programs from formal specifications, and the
construction of a correctness proof hand-in-hand with the program has
been advocated by Dijkstra \cite{Dij76}, Hoare \cite{Ho69}, Gries
\cite{Gr81}, and others.  Central to this method is the a priori
existence of a formal specification which is assumed to 
represent what the user requires.  The task of constructing
such a specification is addressed by the many requirements elicitation
methods appearing in the
literature~\cite{HWT03,HST98,HKLAB98,KKK95,CABetal98,F89,HJL96,MH91,GGH90,MS03},
and is recognized to be the most crucial part of the software life
cycle, as Brooks notes in \cite{Br86}:
\begin{quote}
The hardest single part of building a software system is deciding precisely
what to build. No other part of the conceptual work is so difficult as establishing the detailed technical requirements, including all the interfaces to people, to machines, and to other
software systems. No other part of the work so cripples the resulting system if done wrong. No other part is more difficult to
rectify later.
\end{quote}
%
We aim to ameliorate the difficulty of one crucial part of this
problem: writing a functional specification for a single procedure.
We present a method for the construction of quantified formal specifications
for transformational, terminating, sequential programs. 
Our method automatically constructs a specification, using the results
from a finite number of queries to the user, which the user answers interactively.

Our method relies
on an underlying type theory \tth, which defines the data types over
which the specification is written.
A specification \Sp = \pair{\Pre}{\Post} consists of a precondition, postcondition pair, written in
first order logic.  There is a single input \behi, which is restricted by
the precondition \Pre, and a single
output \beho, which is related to the input by the postcondition \Post.
A specification is satisfied iff the precondition (evaluated on the
input) implies the postcondition (evaluated on the input and the output).
The pair $\beh = \pair{\behi}{\beho}$ is called a behavior, and we write $\beh \sat \Sp$ 
iff $\Pre(\behi) \imp \Post(\behi,\beho)$. We also say that the
behavior is correct w.r.t. the specification.

The problem we address is \emph{specification construction}: to write
a specification which accurately reflects the users
\emph{intentions}. But how are these intentions to be expressed? We
express the users intentions as the answers to a sequence of queries
of the form: ``is the behavior $\behi = \ldots$, $\beho = \ldots$ a
correct behavior?''  Here the $\ldots$ represent the actual variable
values defined by \behi, \beho. Thus, the user identifies those
behaviors which the specification (that is being constructed) must
admit as correct. After a sufficient
number of answers to such queries, our algorithm produces a
specification which accurately reflects the intentions of the user. We
make precise in the sequel this notion of an accurate specification.

In practice, it is easier to produce the precondition and the
postcondition separately, and so we reduce the problem of writing a
specification (a pair of predicates) to the problem of writing a
single predicate, which we call the \emph{formula construction}
problem.

Let \Beh be the set of all possible behaviors. The formula
construction problem is to write a first-order formula 
\Fo that accurately reflects the intentions of the user. 
We take the users intentions to be a partition \Part of \Beh, where
\vtt are the behaviors for which \Fo should evaluate to true, and 
\vff are the behaviors for which \Fo should evaluate to false.
%
%
The user expresses this intention as the answers to a sequence of queries
of the form: ``is the behavior $\behi = \ldots$, $\beho = \ldots$ a
behavior for which \Fo should evaluate to \true? (\ie in \vtt?)
That is, the user
serves as an oracle, and determines for each behavior $\beh =
\pair{\behi}{\beho}$ that our algorithm presents in a query, whether $\beh \in \vtt$
or not.  
\Fo is then constructed as the disjunction of all the $\beh$
that are in $\vtt$, where each \beh is converted to a formula in the
obvious manner.  The problem is that $\Beh$ is infinite in general, so
we need an infinite number of queries to the user, and the resulting
\Fo is infinitely long. We reduce the number of queries, and the
length of \Fo, to finite, by partitioning \Beh into a finite number of
subsets (equivalence classes) such that all the \beh in each class
have the same classification with respect to whether they are in \vtt
or in \vff. We then query the user with one representative from
each class, and construct \Fo
as the disjunction of all cases that the user classifies in \vtt.


Consider a first order theory \gth, \ie a countable set of first order
wff's.  Two elements of $\Beh$ are equivalent with respect to \gth iff they
assign the same truth values to all the formulae in \gth.  Let
$\Beh/\gth$ denote the resulting partition of $\Beh$.
Our goal is to work within a finite partition of \Beh, which results
from using a finite set of wff's. 

Initially, we have available the underlying type theory \tth, which defines the data types over
which the specification is written. This is usually a countable set of wff's.
We next define an equivalence theory \eth, also a countable set of
first order wff's.  The meaning of \eth is that two behaviors in the
same class of \Beh/\eth are ``the same''  with respect to the problem
being specified. For example, if the problem is sorting, then two
input arrays are ``the same'' iff corresponding pairs of elements in
both arrays have the same ordering.
We provide an algorithm to deduce a finite 
version \ethb{b} of \eth, for use in the formula construction
algorithm. Although \ethb{b} is finite, using it directly results, in
general, in impractically long formulae since $\Beh/\ethb{b}$, while
finite, is too large.
To improve the succinctness of the constructed formula \Fo, and also to enable a hierarchical
methodology, we introduce the notion of vocabulary \voc, which is a
finite set of wff's, and which induces a coarser partition $\Beh/\voc$
than $\Beh/\ethb{b}$. \Fo is written using the formulae in \voc. The
number of queries to the user that are needed is bounded by 
$|\Beh/\voc|$, the number of equivalence classes in $\Beh/\voc$.


Now \Fo must express \Part in that \set{\beh \stt \beh \sat \Fo} =
\vtt, \ie \Fo evaluates to true \emph{exactly} on the behaviors in \vtt.
This requires that if \Fo holds for some representative of some class
in $\Beh/\voc$, that it must then hold for all elements in that class.
For this to be possible, no class in 
$\Beh/\voc$ can have elements in both \vtt and in \vff. That is, 
$\Beh/\voc$ must be finer than $\Part$. This property is called
\emph{adequacy} of the vocabulary.  For example, the vocabulary 
\set{a[i] \le a[i+1]} is inadequate to specify sorting, since it does
not enable the expression of the permutation condition: its
equivalence partition is too coarse.  We provide algorithms to both
construct the formula \Fo for \Part given an adequate vocabulary, and
also to check if a vocabulary \voc is adequate, given an equivalence
theory \eth and user answers to \vtt/\vff classification queries.
Our algorithm augments an inadequate vocabulary to make it adequate
by adding some more formulae.

Since each of \tth, \eth, and \voc is written using formulae from the
previous theory, 
$\Beh/\tth$, $\Beh/\eth$, and $\Beh/\voc$ form an increasingly coarser sequence
of partitions of $\Beh$.

In practice $\tth$, $\eth$ and $\voc$ come from (1) the user directly, (2) from
existing specifications and code elements, or (3) from syntax rules restricting
$\tth$ to a finite index theory. In the latter case, $\eth$
and adequacy can be computed while computing \Fo.

\remove{
Section~\ref{s:method} presents technical preliminaries: specifications and behaviors. Section~\ref{s:reduction}
presents our method for constructing a specification.  Sections~\ref{s:ex} and~\ref{s:justify} give two example
applications: linear search and paragraph justification.  
We discuss related work in Section~\ref{s:related}.  We conclude and discuss
future work in Section~\ref{s:conclusion}. Proofs are omitted for lack of space, and can be found in the full version
of the paper, available at \url{http://webfea.fea.aub.edu.lb/fadi/dkwk/doku.php?id=speccheck}.

It is recognized that the task of writing specifications is difficult because it involves formalizing an informal idea
of behavior, given as a set of behaviors (use cases).  Our aim in this paper is to bring some formality and rigour to
this task.
}

\section{Preliminaries: behaviors and specifications}
\label{s:method}
\label{s:prel}


We use (unless otherwise stated) many-sorted first-order logic
\cite[Chapter 4]{End01}.
We use a first order language with both global and local symbols. 
The global symbols are 
 (1) the boolean connectives and equality (=); 
 (2) $n$-ary predicate symbols ($n \ge 0$); and
 (3) $n$-ary function symbols ($n \ge 0$).
The local symbols are the variables.

We are interested in writing specifications for terminating sequential programs, which have a fixed set
$\sv_1,\ldots,\sv_n$ of program variables, which take values from 
universes $\U_1,\ldots,\U_n$, respectively.
A \emph{specification} then consists of a precondition over the initial values of
$\sv_1,\ldots,\sv_n$, and a postcondition that relates the initial and final values of
$\sv_1,\ldots,\sv_n$.

Hence we introduce 
logical variables $\sv_1^i,\ldots,\sv_n^i$ for the initial values of the $\sv_1,\ldots,\sv_n$, and 
logical variables $\sv_1^o,\ldots,\sv_n^o$ for the final values of the $\sv_1,\ldots,\sv_n$.
These will be the only variables in our first order language.
An input state $\asg_i: \tpl{\sv_1^i,\ldots,\sv_n^i} \to \U_1 \times \cdots \times \U_n$ is an
assignment that maps each $\sv_j^i$, $j \in \n$, to a value in its domain, and similarly for an 
output state $\asg_o: \tpl{\sv_1^o,\ldots,\sv_n^o} \to \U_1 \times \cdots \times \U_n$.
A behavior $\asg = (\asg_i, \asg_o)$ is a pair consisting of an input state and an output state.

A formula is interpreted in a many-sorted structure $(I,\asg)$, with a universes $\U_1,\ldots,\U_n$. $I$ provides
the interpretation for the global symbols, and $\asg$ provides the interpretation for the local
symbols, \ie the $\sv_1^i,\ldots,\sv_n^i, \sv_1^o,\ldots,\sv_n^o$.
$I$ provides the usual interpretations of
functions and relations over the integers, etc.  
Let $f$ be a well-formed formula (wff). We write $(I,\asg) \sat f$ iff $f$ is true in the structure
$(I,\asg)$, according to the usual Tarskian semantics.  We usually omit $I$, as it is fixed, and
write $\asg \sat f$. We also write $\asg.f$ for the truth value of $f$ in $(I,\asg)$.
\md{f} denotes \set{ \asg \stt \asg \sat f}, \ie the set of states where $f$ holds.

A \emph{specification} $\Sp = \pair{\Pre}{\Post}$,
 consists of two well-formed formulae: 
 $\Pre$ which represents the precondition, and is restricted to contain only $\sv_1^i,\ldots,\sv_n^i$, and 
 $\Post$, which represents the postcondition. 
A behavior $\asg = \pair{\asg_i}{\asg_o}$ \emph{satisfies} a
specification $\Sp = \pair{\Pre}{\Post}$ iff $\asg.(P \imp Q) = \true$. We
write $\asg \sat \Sp$ in this case, and $\asg \not\sat \Sp$ otherwise.
We also write $\md{\Sp} \df  \set{ \asg \stt \asg \sat \Sp }$.

Let \Beh be the set of behaviors. We partition \Beh into:
\be 
\i $\good$, the set of good (positive) behaviors: the precondition
holds before and the postcondition holds after, \ie 
$\beh.P = \true$ and $\beh.Q = \true$ for all $\beh \in \good$;
\i $\bad$, the set of bad (negative) behaviors: the precondition holds
before and the postcondition does not hold after, \ie 
$\beh.P = \true$ and $\beh.Q = \false$ for all $\beh \in \bad$; and
\i $\dc$, the set of don't care behaviors: the precondition does not
hold before, and the postcondition can be either true or false after,
\ie $\beh.P = \false$ for all $\beh \in \dc$.
\ee
%
A partition $(\good, \bad, \dc)$ of $\Beh$ is \emph{feasible} iff 
(1) for every input state $\asg_i$, there do not exist two output states $\asg_o, \asg_{o'}$ such
that  $(\asg_i,\asg_o) \in \good \un \bad$ and $(\asg_i,\asg_{o'}) \in \dc)$; and
(2) for every input state $\asg_i$ there exists an output state $\asg_o$ such that $(\asg_i,\asg_o) \in \good \un \dc)$. 
Clause~(1) means that the precondition is not both true $((\asg_i,\asg_o) \in \good \un \bad)$ and false $((\asg_i,\asg_{o'}) \in \dc)$ when
evaluated on input $\asg_i$. Clause~(2) means that for every input there is at least one acceptable output.
In the sequel, we consider only feasible partitions of $\Beh$. We assume 
that the developer can reliably classify given behaviors as good,
bad, and dontCare behaviors.


%

\section{The Specification and Formula Construction Problems}
\label{s:problem}

\begin{definition}[Specification construction problem]
\label{def:spec-construction}
Let $(\good, \bad, \dc)$ be a feasible partition of $\Beh$. 
The \emph{specification construction problem} is to find a
specification \Sp such that $\md{\Sp} = \good \un \dc$.
\end{definition}
We say that such an \Sp is \emph{accurate} with respect to $(\good,\bad,\dc)$.

\begin{definition}[Formula construction problem]
\label{def:formula-construction}
Let $(\vtt,\vff)$ be a partition of $\Beh$.
The \emph{formula construction problem} is to find a wff \Fo such that \md{\Fo} = \vtt.
\end{definition}
We say that such a \Fo is \emph{accurate} with respect to $(\vtt,\vff)$.
We reduce specification construction to formula construction, as
follows.  Construct a formula $\Pre$ that is accurate w.r.t. $( \good \un
\bad, \dc )$.  Also construct a formula $\Post$ that is accurate
w.r.t. $(\good \un \phi, \bad \un \psi)$, where $(\phi, \psi)$ is an
arbitrary partition of \dc, which can be chosen for convenience of
expressing $\Post$.
From the definitions of $\asg \sat \Sp$ and $\md{\Sp}$ given above, we obtain 
$\md{\Sp} = \good \un \dc$, and so $\Sp$ is accurate with respect to $(\good,\bad,\dc)$.


\section{The Formula Construction Algorithm}
\label{s:reduction}

Let $F$ be a set of first order wff's.  $V: F \to \Bool$ is a \emph{valuation of $F$}, 
\ie a mapping that assigns to each $f \in F$ a truth-value.
Write $F \mapsto \Bool$ for the set of valuations of $F$.
Define $\for{V} \df (\land f \in F : f \ev V.f)$, \ie $\for{V}$ is the formula which asserts that each $f \in F$ has the
truth value assigned to it by $V$.
$\for{V}$ can be infinitely long, \ie it is a formula of the
infinitary logic $\IL$ \cite{Be12}. 
Define $\md{V} \df \set{\asg \stt \asg \in \Beh \land (\land f \in F: \asg.f = V.f)}$, 
\ie $\md{V}$ is the set of all
behaviors that assign the same values to the formulae in $F$ that $V$ does.
Note that $\md{V} = \md{\for{V}}$.
$F_{\sim} = \set{\tpl{\asg,\asg'} \stt (\land f \in F: \asg.f = \asg'.f)}$ is the equivalence relation on $\Beh$ that
considers two elements equivalent iff they assign the same values to the formulae in $F$.
Thus $\Beh/F_{\sim} = \set{ \md{V} \stt V \in F \mapsto \Bool}$ is a partition of $\Beh$.
We assume the standard definitions for one partition of $\Beh$ being finer (coarser) than another, and write 
$\Beh/E \le \Beh/E'$ when $\Beh/E$ is finer than $\Beh/E'$.

\subsection{Type theory, equivalence theory, and vocabulary}

Our solution to the formula construction problem rests on three
foundations: (1) the use of an underlying \emph{type theory} \tth,
which defines the domains of all the free variables, and also all the
operators over these variables; and (2) the use of an
\emph{equivalence theory} \eps, which defines an equivalence relation
over $\Beh$; and (3) the use of a \emph{vocabulary} \voc for
constructing our formula \Fo.  All of these are sets of first order
wff's.



\subsubsection{Type theory}
The type theory \tth represents the ``finest granularity of expression''
that we have. It provides the basic axioms for all the free variables, \eg integer scalars, array indices,
arrays, etc. See for example, \cite{Brdly06} for an example array
theory, and \cite{Furia2010} for an example theory of sequences.
In practice, the type theory is supported by the SMT solver.

\subsubsection{Equivalence theory}
The user is required to provide the equivalence theory \eps as the 
\empb{primitive formalization step}. 
Two behaviors that induce the same valuations of all formulae in $\eth$ are considered equivalent:
$\asg \ev_\eth \asg' \df (\fa f \in \eth : \asg.f = \asg'.f)$, which can also be written
$\asg \ev_\eth \asg' \df (\ex \Ve: \asg \in \md{\Ve} \land \asg' \in \md{\Ve})$.
Thus, \eth represents the users opinion of which parts of $\Beh$ will
be considered equivalent 
with respect to the problem being specified. 
We illustrate with two examples:
\be

\item search of an array $a$ between indices $\l$ and $r$ inclusive:
\be
\i \set{|a| = n \mbox{\ for all $n > 0$}}
\i $\l = r$, $\l < r$ 
\i $\l \ge 0$, $\l \le |a|-1$,\\ \set{\l = c \mbox{\ for all $c$ such that $0 \le c < |a|$}}
\i $r \ge 0$, $r \le |a|-1$,\\ \set{r = c \mbox{\ for all $c$ such that $0 \le c < |a|$}}

\i \set{a[i] = e \mbox{\ for all $i$ such that $0 \le i < |a|$}}
\i \set{rv = c \mbox{\ for all $c$ such that $0 \le c < |a|$}}
\i $rv = -1$
\ee


That is, values for the left bound $\l$, are equivalent iff either they are equal or they
are both out of bounds in the same manner (too low, too
high). Likewise for the right bound $r$. Values for the array $a$ are equivalent
iff they are the same size and corresponding elements are equivalent w.r.t. matching the search
expression $e$. Values for the return index $rv$ are equivalent if
they are both -1, or they both indicate the same position in the array.

\item sort an array $a$ : 
\be
\i \set{|a| = n \mbox{\ for all $n > 0$}}
\i \set{a[i] < a[j] \mbox{\ for all $i,j$ such that $0 \le i < j < |a|$}}
\i \set{a[i] = a[j] \mbox{\ for all $i,j$ such that $0 \le i < j < |a|$}}
\ee

That is, values for array $a$ are equivalent iff they are the same
size and corresponding pairs of elements have the same order and equality 
relationship. 

\ee

We express an equivalence theory using both single formulae, \eg
$\l = r$, and sets of formulae, \eg 
\set{\l = c \mbox{\ for all $c$ such that $0 \le c < |a|$}}.
For now we assume that \eth is a finite set of wff's, and so, \eg, we
restrict $|a|$ to a finite value. We show in Section~\ref{sec:finite}
how to deal with equivalence theories consisting of a countable set of
wff's.


\remove{
The above examples are expressed using metasyntax, since the set of formulae is infinite, due to the presence of arrays,
whose size is unbounded, in general.
Let \ethm be the metasyntactic representation of \eth, \ie \eth results from \emph{expanding} \ethm. 
\ethm consists of single formula, such as $\l < r$, and sets of formulae, such as 
\set{\l = c \mbox{\ for all $c$ such that $0 \le c < |a|$}}, where any predicate can be used for the range. Formalizing
the metasyntax is straightforward. Expansion replaces each construct \set{\mbox{$f$ such that $R$}} by
a set of formulae, one for each satisfying assignment of the range predicate $R$, in the obvious manner.

Define \ethmb{b} to be the metasyntax obtained by textual substitution of $b$ for all array sizes $|a|$ in the range
predicate of every set expression of \ethm by $b$, where $b > 0$. 
All other occurrences of $|a|$ (\eg $\l \le |a|-1$) are not replaced by $b$.
Let \ethb{b} be the result of expanding \ethmb{b}. 
For example, if \ethm is the metasyntax for search given above, then \ethb{2} contains \set{\l=0, \l=1} while 
\ethb{3} contains \set{\l=0, \l=1, \l=2}.
We require:
\newsavebox{\finiteEquiv}
\sbox{\finiteEquiv}{Finite-Equiv}
\bleqn{\usebox{\finiteEquiv}}
\parbox{4.5in}{
\ind $\fa b > 0 : \mbox{$\ethb{b}$ is a finite set}$ and $\ethb{b} \sub \ethb{b+1}$.
}
\eleqn
For example, the metasyntax given above satisfies \usebox{\finiteEquiv}, while the 
metasyntax \set{\l = c \mbox{\ for all $c$ such that $0 \le c$}} does not, as it generates $\l=0, \l = 1, \ldots$.
The requirement $\ethb{b} \sub \ethb{b+1}$ can be satisfied by making all the range predicates in set expressions
monotonic in all array sizes.
}

We argue that equivalence is a notion that is intuitively well
understood informally, and is moreover relatively 
easier to formalize than a complete specification. We justify this
claim by a series of examples: since the claim is inherently informal,
it can be supported only by empirical evidence.


%
\eth provides an initial set of wff's
that are ``building blocks'' for the formula \Fo that we are constructing. We cannot use \tth directly for this, since
the number of formulae that must be considered is infinite, in general, and so we have computability limitations.
Since \tth is our basic vocabulary, all formulae of \eps are
written using formulae in $\tau$. Hence we immediately obtain:

\begin{proposition}
$\Beh/\tau \le \Beh/\eth$
\end{proposition}

To be able to express \Part using \eth, we assume the following axiom:
\newsavebox{\AxiomEquiv}
\sbox{\AxiomEquiv}{Axiom-E}
\bleqn{\usebox{\AxiomEquiv}}
$\Beh/\eth \le \Part$
\eleqn
That is, we consider only partitions \Part that are coarser that $\Beh/\eps$. 
This is reasonable, since the number of partitions of $\Beh$ is usually uncountable (when the input and output variables
are boolean, string, and integer), while the number of formulae that we can write is countable (assuming that the
number of variables, and function and predicate symbols, is countable). Hence some restriction on the partitions that
can be expressed is required.
Also, \eps defines the pairs of valuations $\asg, \asg'$ that the user considers equivalent. Thus any predicate that the user wishes
to define must take the same values on $\asg$ and $\asg'$. So, \Fo  must correspond to partitions
coarser than $\Beh/\eth$. This is reasonable, since \eps is the first formalization step, and so defines in effect the
formal elements from which \Fo is written.

We emphasize that the \emph{only} requirement on \eth is \usebox{\AxiomEquiv}, and so writing \eth is easier than
writing a full and accurate specification from scratch.


\subsubsection{Vocabulary}

By \usebox{\AxiomEquiv}, each $\md{V_\eth}$ is wholly contained in either \vtt or in \vff. Hence the union of all $V_\eth$ that
are contained in \vtt is exactly \vtt: $\vtt = (\union V_\eth : \md{V_\eth} \sub \vtt : \md{V_\eth})$. Hence, the
disjunction of all the formulae \for{V_\eps} corresponding to $V_\eth$ that
are contained in \vtt yields a formula which is true at all elements of \vtt and false outside of \vtt. That is, 
 $\Fo = (\OR V_\eth : \md{V_\eth} \sub \vtt : \for{V_\eth})$.
is a tentative solution to the formula construction problem since \md{f} = \vtt.
However, in practice, this solution is far too verbose to be useful, since the number of equivalence classes in $\Beh/\eth$ is far too large, each
such class consisting of all valuations that the user considers ``equivalent'' w.r.t. the specific problem being
solved. The vocabulary $\nu$ introduces the coarser building blocks needed to write \Fo succinctly. 
Its formulae are constructed from those of \eps, and so we have:
\begin{proposition}
\label{prop:equiv-finer-voc}
$\Beh/\eth \le \Beh/\voc$
\end{proposition}
by construction of \voc. To be able to express \Part using \voc, we require
%
\newsavebox{\AdequacyCondition}
\sbox{\AdequacyCondition}{(Ad)}
\bleqn{\usebox{\AdequacyCondition}}
$\Beh/\voc \le \Part$
\eleqn
that is, $(\fa \Vv : \md{\Vv} \sub \vtt \lor \md{\Vv} \sub \vff)$.
We call such a \voc \emph{adequate}. Unlike the situation for \eps, we cannot take $\Beh/\voc \le \Part$ as an axiom,
since \voc can contain arbitrarily coarse formulae, \ie formulae $f$ with large \md{f}. In practice, we wish to use the
coarsest formulae possible, since this will give the most succinct expression of \Part. So often, $\Beh/\voc$ will be too
coarse, violating $\Beh/\voc \le \Part$, and will have to be corrected. 
A heuristic for writing an adequate vocabulary \voc is that \voc should contain wff's for every concept in the
initial informal natural language description of the problem, \eg for both ordering and permutation in the
case of array sorting.

We give an algorithm for checking adequacy of \voc and correcting an inadequate \voc in Section~\ref{sec:adequacy}. 
This process of approximating  the coarsest adequate vocabulary can be viewed as an \emph{abstract
interpretation} \cite{CC77} problem. 

Within this section, we assume that \voc is adequate.
Hence each $\md{V_\voc}$ is wholly contained in either \vtt or in \vff. Hence the union of all $V_\voc$ that
are contained in \vtt is exactly \vtt: $\vtt = (\union V_\voc : \md{V_\voc} \sub \vtt : \md{V_\voc})$.
We can thus improve our tentative solution to $\Fo = (\OR V_\voc : \md{V_\voc} \sub \vtt : \for{V_\voc})$.
With a coarse enough \voc, this will generate a succinct \Fo.

\subsection{The formula construction algorithm}

Our algorithm evaluates $\Fo = (\OR V_\voc : \md{V_\voc} \sub \vtt : \for{V_\voc})$, \ie it constructs \Fo as the
disjunction of the formulae \for{V_\nu} for each $V_\voc$ contained in \vtt. 
There are $2^{|\voc|}$ different assignments $V_\voc$. 
We start with \Fo set to false, and we loop through these. For each $V_\voc$, we 
submit $\for{V_\voc}$ to a Satisfiability-modulo-theory (SMT) solver, \eg Z3~\cite{Z309}. 
An SMT solver takes as input a formula in a defined theory under first order logic.
There are three possible outcomes:
(1) the SMT solver exhausts its computational resources before determining if $\for{V_\voc}$ is satisfiable,
(2) the SMT solver returns a satisfying assignment for $\for{V_\voc}$, and 
(3) the SMT solver returns that $\for{V_\voc}$ is unsatisfiable.

In case (1), our algorithm for constructing \Fo terminates with failure. The developer can use the feedback from the
failed attempt, such as the unsat core, to try to simplify the problem, \eg by modifying \voc, and then re-attempting.
In case (2), 
a satisfying assignment $\asg$ is a partial assignment to the variables in $\for{V_\voc}$; \ie a partial
assignment to the variables in $vars(\voc)$. 
The remaining variables in $vars(\voc)$ can be assigned arbitrarily without affecting the satisfiability of $\for{V_\voc}$.
Since we now have a value for each variable in $vars(\voc)$, we can interpret $\asg$ (augmented with the arbitrary
assignments) as an element of $\Beh$.
We present $\asg$ to the developer, who determines whether $\asg \in \vtt$ or $\asg \in \vff$.  
Thus we really require the developer to classify an assignment $\asg$ as either 
``in the set of assignments for which the formula should be true'' or 
``in the set of assignments for which the formula should be false'', 
and we assume that this classification is accurate. 
If the developer responds ``in \vtt'', then we conclude, by \usebox{\AdequacyCondition}, that $\md{V_\voc} \sub \vtt$.
Hence we update \Fo by disjoining $\for{V_\voc}$ to it,  as indicated by the pseudocode line  
$\Fo := \Fo \cat ``\lor\mbox{''} \cat \for{V_\voc}$ in Figure~\ref{a:reduction},
where $\cat$ denotes string concatenation, \ie we are constructing the text of the formula $\Fo$ as a concatenation of
disjuncts.
By construction, each disjunct $\for{V_\voc}$ is a conjunction of literals. Thus, 
$\Fo$ can be simplified each time a disjunct
is added, using sum of products simplification, or all at once after 
\constFor\ terminates.
In case (3), we conclude, by \usebox{\AdequacyCondition}, that $\md{V_\voc} \sub \vff$, so we do not alter \Fo.

We iterate the above for every valuation $V_\voc: \voc \to \Bool$ and so we compute
$\Fo$ as the disjunction of the $\for{V_\voc}$ such that $\asg$ is in $\vtt$.  We annotate the pseudocode in
Figure~\ref{a:reduction} with a loop invariant and some Hoare-style annotations.  We use an auxiliary variable $\vphi$,
which records the valuations $V_\voc$ that have been processed so far. The correctness of these annotations is
self-evident from the pseudocode and the assumption of an adequate vocabulary \voc.
Figure~\ref{a:reduction} presents algorithm \constFor($\voc, \vtt, \vff$) which takes as input a
partition \Part of $\Beh$ and an adequate vocabulary \voc, and returns a formula \Fo such that $\md{\Fo} = \vtt$.
Theorem~\ref{thm:accurate} below follows immediately from the previous discussion.

\begin{figure*}[t!]
\center{
\fbox{\parbox{\textwidth}{
\begin{tabbing}
mm\=mm\=mm\=mm\=mm\=mm\= \kill
\constFor($\voc, \vtt, \vff$)\\[1ex]
\asrt{\mbox{Precondition: (\vtt,\vff) partitions $\Beh$ and $\Beh/\voc \le \Part$ } }\\
$\Fo := \false$; $\vphi := \Vv \mapsto \Bool$\\
\asrt{\mathrm{Invariant}: \Fo \ev (\OR \Vv : \Vv \not\in \vphi \land \md{\Vv} \sub \vtt : \for{\Vv})}\\
\WHILE $\vphi \ne \emptyset$\\
 \>select some valuation $\Vv \in \vphi$;\\
  \>$\vphi := \vphi - \Vv$;\\
  \>submit $\for{\Vv}$ to an SMT solver;\\
  \>\IF the solver succeeds \THEN\\
  \>  \>\IF $\for{\Vv}$ is satisfiable \THEN\\
  \>  \>  \>let $\asg_v$ be the returned satisfying assignment;\\
  \>  \>  \>query the developer: is $\asg_v$ in \vtt or in \vff?\\ 
  \>  \>  \>\IF developer answers $\asg_v \in \vtt$ \THEN\\
  \>  \>  \>   \>$\Fo := \Fo \cat ``\lor\mbox{''} \cat  \for{\Vv}$; \cmnt can simplify \Fo to improve succinctness\\
  \>  \>  \>\ELSE\\
 \>  \>  \>   \>$\mathit{skip}$;    \cmnt can use partial assignment to reduce $\vphi$ (Section~\ref{sec:partial-assig}).\\
  \>  \>  \>\ENDIF \\
  \>  \>\ELSE \>  \cmnt solver returned unsat \\
  \>  \>  \>let $\mathit{unsat} \subseteq \Vv \mapsto \Bool$ be the unsat core valuations;\\
  \>  \>  \>$\vphi := \vphi - \mathit{unsat}$; \cmnt unsat core reduction (Section~\ref{sec:unsat-core}) \\
  \>  \>\ENDIF \\ 
  \>\ELSE \RETURN(``failure'') \cmnt return with failure since solver cannot answer query\\
  \>\ENDIF\\
\ENDWHILE\\
\asrt{\mbox{Postcondition:}\ \Fo \ev (\OR V_\voc : \md{V_\voc} \sub \vtt : \for{V_\voc})}\\
\RETURN$(\Fo)$\\
\end{tabbing}
\vspace{-1em}
}}}
\caption{\constFor($\voc, \vtt, \vff$)}
\label{a:reduction}
\end{figure*}



\begin{theorem}[Correctness of \constFor]
\label{thm:accurate}
Assume that (1) $voc$ is \emph{adequate} for $(\vtt, \vff)$ and (2) that 
no invocation of the SMT solver by \constFor\ fails, and (3) the developer responds accurately to all
queries.
Then \constFor\ returns formula $\Fo$ such that $\md{\Fo} = \vtt$.
\end{theorem}

\subsection{Decidability and Complexity}
\label{sec:complexity}

\constFor\  may fail to generate a formula if the SMT solver fails on any call.
We are therefore interested in subclasses of first order logic where success is guaranteed.
For example, when each of the $\for{\Vv}$ formulae belongs to a class of formulae 
solvable in a finite domain,
such as equality, monadic, and quantifier free theories~\cite{Akrmn54}, and 
array theories 
with one quantifier alternation under syntactic restrictions~\cite{Brdly06,Furia2010}
that can be reduced to the combined theory of equality with 
uninterpreted functions (EUF).
Such theories are enough to express specifications such as sortedness and 
injectivity. 

The running time of \constFor\ is at most $2^{|\voc|}$ calls to the SMT solver, since
everything else is straight-line code. Following are three improvements that in practice give
us significant reductions in the number of calls to the solver.


We discuss next optimizations that further reduce the number of
user queries needed.

\subsubsection{Unsat-core elimination} 
\label{sec:unsat-core}
When \for{\Vv} is found to be unsatisfiable, we obtain the 
unsat core from the SMT solver, and eliminate from consideration all \Vv that are extensions of the
unsat core, since all of these will be unsatisfiable.

\subsubsection{Partial-assignment elimination}
\label{sec:partial-assig}
The user can eliminate many valuations in one step as follows. 
When the user deems a presented assignment to be in \vff, the user can select a subset
of the variables assigned as the real reason for the choice.  
The partial assignment selected by the user may leave some
of the subformulas in $\voc$ not evaluated to a truth value.  
For example, consider $\voc=\{C_1,C_2,C_3\}$ and consider
a \vff assignment $l=-1,r=1$ where the user selects $l=-1$ 
as the reason for the \vff decision.  
The partial assignment selected by the user evaluates $C_1$ 
and $C_2$ to a truth value, but leaves $C_3$ dependent 
on $r$.  
We learn that the valuation corresponding to 
$\neg C_1 \land \neg C_2 \land \neg C_3$ is also a 
\vff without further querying the SMT
solver and the user.  
We use the partial assignment selection by the user to reduce the number of valuations that we
consider. We can apply this idea to \vtt also, \ie partial-assignment inclusion.

\subsubsection{Hierarchical construction of vocabularies}
\label{sec:hierarchical-vocab}
We expect $|\voc|$ to be small in many cases, as it is the number of formulae used to construct a formula at the next
level. In practice, we can keep the running time reasonable by constructing the vocabulary hierarchically, and
structuring the levels of the hierarchy so that a formula is
not constructed out of too many lower-level components. Such a long formula is difficult to write correctly using
informal techniques, and so the methodological practices that make our method efficient are those that are a good idea
in any case.

\section{Adequacy of the vocabulary}
\label{sec:adequacy}

We wish to verify that $\voc$ is adequate: $\Beh/\voc \le \Part$.
Proposition~\ref{prop:equiv-finer-voc} gives us $\Beh/\eth \le \Beh/\voc$.
Hence each \md{\Vv} is a union of some \md{\Ve}. 
\usebox{\AxiomEquiv} gives us $\Beh/\eth \le \Part$, that is 
all the elements of each  equivalence class \md{\Ve} are in the same partition (\vtt or \vff) of $\Beh$, 
and so it suffices to check a single representative of each \md{\Ve}.
We check that the \md{\Ve} that make up \md{\Vv} are either all contained in \vtt, or all contained in \vff. 
This implies that \md{\Vv} is contained in \vtt or is contained in \vff, \ie $\Beh/\voc \le \Part$.

We present an algorithm to check adequacy when \eth and \voc are both
finite. We discuss in the next section how to handle general \eth and \voc.
The check is implemented as follows. We iterate over all the \md{\Ve}, and for each we find a representative $\asg_{\Ve} \in \md{\Ve}$
by invoking an SMT solver on $\for{\Ve}$. If $\for{\Ve}$ is not satisfiable, then it defines an empty partition of
$\Beh/\eth$ (which is certainly possible) and so we do nothing. Otherwise, 
a satisfying assignment gives a $\asg_{\Ve} \in \md{\Ve}$. 
We query the user as to whether $\asg_{\Ve}$ is in \vtt or in \vff, and record the result.

We then iterate over all the \Vv, and for each we iterate over all the \md{\Ve},
checking if $\asg_{\Ve} \sat \for{\Vv}$. 
If so, then $\md{\Ve} \sub \md{\Vv}$ by the above discussion, since either all elements of 
$\md{\Ve}$ are in  $\md{\Vv}$ (and so satisfy $\for{\Vv}$), or none are (in which case none satisfy $\for{\Vv}$).
We look up the classification of $\asg_{\Ve}$ (in \vtt or in \vff). If $\asg_{\Ve} \in \vtt$ then we know that \Vv
intersects \vtt, since $\md{\Ve} \sub \md{\Vv}$. Likewise if $\asg_{\Ve} \in \vff$ then we know that \Vv
intersects \vff. A \Vv that intersects both \vtt and \vff is a cause of inadequacy of \voc, since it causes 
$\Beh/\voc \le \Part$ to be violated. We correct this by adding the ``correction formula''
$(\OR \Ve : \asg_{\Ve} \sat \for{\Vv} \land side[\Ve] = ``\vtt\mbox{''} : \for{\Ve})$ to \voc. This  
splits \md{\Vv} into $\md{\Vv} \ints \vtt$ and $\md{\Vv} \ints \vff$. We compute all such needed correction formulae and
store them in an array \cfor{} which our algorithm, given in Figure~\ref{a:adequacy}, returns.


\begin{figure*}[t!]
\center{
\fbox{\parbox{\textwidth}{
\begin{tabbing}
mm\=mm\=mm\=mm\=mm\=mm\= \kill
\mkAdeq($\voc, \eth, \vtt, \vff$)\\[1ex] 

\asrt{\mbox{Precondition: \voc and \eth are finite}}\\

\FOREACH valuation $\Ve: \eth \to \Bool$\\
   \>submit $\for{\Ve}$ to an SMT solver;\\
   \>\IF the solver cannot answer the query\\
   \>   \>terminate with failure;\\
   \>\ELSF  the solver succeeds and returns that $\for{\Ve}$ is not satisfiable \THEN;\\
   \>   \>\skipp;\\
   \>\ELSE  \\
   \>   \>let $\asg_{\Ve} \sat \for{\Ve}$ be satisfying assignment returned by solver; \cmnt $\asg \in \md{\Ve}$\\
   \>   \>query the developer: is $\asg$ in \vtt or in \vff?\\ 
   \>   \>$side[\Ve] :=$ answer of developer   \cmnt either ``\vtt'' or ``\vff''. Store in $side[\Ve]$\\
\ENDFOR\\[1ex]

\FOREACH valuation $\Vv: \voc \to \Bool$\\
  \>\FOREACH valuation $\Ve: \eth \to \Bool$\\
  \>   \>\IF $\asg_{\Ve} \sat \for{\Vv}$                 \cmnt $\md{\Ve} \sub \md{\Vv}$\\
  \>   \>   \>$sides[\Vv] := sides[\Vv] \un side[\Ve]$  \cmnt \Vv intersects $side[\Ve]$\\
  \>\ENDFOR                   \cmnt Accumulate in $sides[\Vv]$\\\\
\ENDFOR\\[1ex]

\FOREACH valuation $\Vv: \voc \to \Bool$\\
   \>\IF $sides[Vv] = \set{\vtt,\vff}$  \cmnt \Vv intersects \vtt and \vff\\
   \>   \>$\cfor{\Vv} := (\OR \Ve : \asg_{\Ve} \sat \for{\Vv} \land side[\Ve] = \vtt : \for{\Ve})$\\
   \>\ELSE\\ 
   \>   \>$\cfor{\Vv} := \true$\\
   \>\ENDIF\\
\ENDFOR\\
\RETURN(\cfor{});\\
\end{tabbing}
\vspace{-1em}
}}}
\caption{\mkAdeq($\voc, \eth, \vtt, \vff$)}
\label{a:adequacy}
\end{figure*}

It follows from the above discussion that when all the correction formulae given by \cfor{} are added to \voc, the
result is an adequate vocabulary.

\begin{theorem}[Correctness of \mkAdeq]
\label{thm:MakeAdequate-correct}
Let \mkAdeq($\voc, \eth, \vtt, \vff$) return the array \cfor{} of correction formulae. Then 
$\voc \un (\un \Vv \in \voc \mapsto \Bool : \cfor{\Vv})$ is an adequate vocabulary.
\end{theorem}

\section{Finiteness and Decidability Considerations}
\label{sec:finite}
We will discuss the 
the results of \cite{Brdly06}, namely
a decidable fragment of first order logic that can express 
some properties of arrays in Section~\ref{sec:quantformconst}.
Here we present our reduction
of \eth to \ethb{b}, a finite version of \eth where arrays have size $b$.


The algorithms given above assume that \eth and \voc are finite sets
of wff's, since otherwise the number
of equivalence classes is uncountable, in general. To remove this
restriction, we first formalize the notation in which we
express an equivalence theory. An element of $\Beh$ defines values for some
scalar variables $\bar{z}$ (\eg booleans and integers) 
and some arrays $\bar{a}$.
For ease of exposition, we assume that there is
exactly one array $a$. It is straightforward to remove this restriction.
Let $\bar{i}$ be a set of ``dummy'' variables, which we
use to index $a$.

\bd[Equivalence Theory Syntax] \label{def:equiv-syntax}
An equivalence theory \eth consists of a finite number of \emph{scalar formulae}
$g_1(\bar{z}, |a|), \ldots, g_m(\bar{z}, |a|)$, and a finite number of \emph{indexed
formula set expressions} \set{ f_1(\bar{z}, a, \bar{i}) \stt r_1(|a|, \bar{i}) },$\ldots$,
\set{ f_n(\bar{z}, a, \bar{i}) \stt r_n(|a|, \bar{i}) }.

The range predicate $r(|a|, \bar{i})$ must
be monotonic in $|a|$: for $b' > b$, 
$\set{\bar{v} \stt r(\bar{v}, b)} \sub \set{\bar{v} \stt r(\bar{v}, b')}$.
\ed
As indicated, a scalar formula can refer to the scalar variables $\bar{z}$
and to the size $|a|$ of array $a$. An indexed formula $f(\bar{z}, a, \bar{i})$
can refer to the $\bar{z}$, and to elements of $a$ by using any of $\bar{i}$ as
an index.  The range predicate $r(|a|, \bar{i})$ can refer to $\bar{i}$ and
$|a|$.

\bd[Bounded Equivalence Theory \ethb{b}] \label{def:equiv-bounded}
For $b > 0$ and equivalence theory \eth, the equivalence theory with bound $b$,
\ethb{b}, is the set of wffs
$\set{g_1(\bar{z}, b), \ldots, g_m(\bar{z}, b)}$ $\un$
\set{ f_1(\bar{z}, a, \bar{i}) \stt r_1(b, \bar{i}) }, $\un \ldots \un$,
\set{ f_n(\bar{z}, a, \bar{i}) \stt r_n(b, \bar{i}) }.
Each formula set expression \set{ f(\bar{z}, a, \bar{i}) \stt r(b, \bar{i}) }
denotes the set of formulae consisting of $f(\bar{z}, a, \bar{v})$ for each $\bar{v}$ such
that $r(b,\bar{v})$. 
\ed
Example: search of an array $a$ between indices $\l$ and $r$ inclusive:
\be 
\i $\l = r$, $\l < r$ 
\i $\l \ge 0$, $\l \le |a|-1$,\\ \set{\l = c \mbox{\ for all $c$ such that $0 \le c < |a|$}}
\i $r \ge 0$, $r \le |a|-1$,\\ \set{r = c \mbox{\ for all $c$ such that $0 \le c < |a|$}}
\i \set{a[i] = e \mbox{\ for all $i$ such that $0 \le i < |a|$}}
\ee
For $|a| = 5$, we obtain:
\be
\i $\l = r$, $\l < r$ 
\i $\l \ge 0$, $\l \le 4$, $\l=0$, $\l=1$, $\l=2$, $\l=3$, $\l=4$ 
\i $r \ge 0$, $r \le 4$, $r=0$, $r=1$, $r=2$, $r=3$, $r=4$ 
\i  $a[0]=e$, $a[1]=e$, $a[2]=e$, $a[3]=e$, $a[4]=e$ 
\ee
Here $c$ and $i$ are the dummies.

We wish to find a ``threshold'' \th such that we can execute our algorithms
using \ethb{\th} instead of \eth. Since \eth is the union of \ethb{b} for all $b
> 0$, we must show how every \ethb{b} can be ``represented'' in \ethb{\th}.
We require that every satisfiable valuation $\Veb{b}$ in \ethb{b} 
have a representative valuation in \ethb{\th}. 
We will then process this representative, rather than $\Veb{b}$. If we can do
this for all valuations $\Veb{b}$ for all $b > \th$, then we can replace reasoning
about the infinite theory \eth with reasoning about the finite theory \ethb{\th}.

Given $\th, b$ such that $\th < b$, we define the mapping 
$M_{\th b}: \ethb{\th} \mapsto \ethb{b}$ as follows.
For $j = 1,\ldots,m$, $g_j(\bar{z}, \th)$ maps to $g_j(\bar{z}, b)$.
For $k = 1,\ldots,m$, 
$f_k(\bar{z}, a, \bar{v})$ maps to $f_k(\bar{z}, a, \bar{v})$
for each $\bar{v}$ such that $r_k(\th, \bar{v})$ holds. 
Note that $r_k(b, \bar{v})$ also holds, by monotonicity of range predicates.

For each valuation \Veb{b}, we define the projection onto \ethb{\th}, 
$\Veb{b} \pj \ethb{\th}$: for every $f \in \ethb{\th}$, 
$\Veb{b} \pj \ethb{\th}(f) =  \Veb{b} (M_{\th b}(f))$. That is, we evaluate a
formula $f$ of \ethb{\th} in $\Veb{b} \pj \ethb{\th}$ by mapping it to 
$\Veb{b}$ using $M_{\th b}$, and then applying $\Veb{b}$.

We will use $\Veb{b} \pj \ethb{\th}$ as the representative of \Veb{b}.
For our algorithms to work correctly under this mapping, we require, for some
$\th$ and all $b > \th$:
\bn
\item \label{project-val-satis}
If $\Veb{b}$ is satisfiable,  then so is $\Veb{b} \pj \ethb{\th}$. That
  is, if $\md{\Veb{b}} \ne \emptyset$, then $\md{\Veb{b} \pj \ethb{\th}} \ne
  \emptyset$.

\item \label{project-val-user}
For all $\asg_b \in \md{\Veb{b}}$,
          $\asg_{\th} \in \md{\Veb{b} \pj \ethb{\th}}$, 
    the user classifies $\asg_b$ and $\asg_{\th}$ in the same way, \ie both in
    \vtt or both in \vff.
\en

Clause~\ref{project-val-satis} can be checked mechanically by 
submitting it to a SMT solver.
Our first attempt to write Clause~\ref{project-val-satis} as a first
order wff is:
$$\ex \th \fa b > \th: (\satf \for{\Veb{b}}) \ \imp\ (\satf \for{\Veb{b} \pj \ethb{\th}}),$$
where we render each occurrence of $\satf$ using existential quantification 
over boolean variables, \ie bits. 

However, the formulae in \ethb{b} depend on $b$, which presents a problem:
$(\satf \for{\Veb{b}})$ is a wff which depends on $b$, so that different $b$ give
different formulae. Thus, we have to check an infinite set of wff's, one for
each $b$. We deal with this by verifying a single formula which implies each of
these wff's.

Define $\for{\Veb{b}} \df \for{\Veb{b}^s} \land \for{\Veb{b}^i}$, where 
$\for{\Veb{b}^s}$ is the assignment to the scalar formulae in $\ethb{b}$, and
$\for{\Veb{b}^i}$ is the assignment to the indexed formulae in $\ethb{b}$.
Likewise define 
$\for{\Veb{\th}} \df \for{\Veb{\th}^s} \land \for{\Veb{\th}^i}$,
where $\Veb{\th} \df \Veb{b} \pj \ethb{\th}$.
We wish to check
$$(\satf\for{\Veb{b}^s} \land \for{\Veb{b}^i}) \imp
(\satf \for{\Veb{\th}^s} \land \for{\Veb{\th}^i}).$$

By monotonicity of range predicates, we have $\ethb{\th}^i \sub \ethb{b}^i$,
where $\ethb{\th}^i, \ethb{b}^i$ are the subsets of $\ethb{\th}, \ethb{b}$
respectively, consisting of the indexed formulae. Hence 
$\for{\Veb{b}^i} \imp \for{\Veb{\th}^i}$ is logically valid. 
So
$$(\satf \for{\Veb{b}^s} \land \for{\Veb{b}^i}) \ \imp\ 
    (\satf \for{\Veb{b}^s} \land \for{\Veb{\th}^i})$$ is also logically valid.
Hence it suffices to check
$$(\satf \for{\Veb{b}^s} \land \for{\Veb{\th}^i}) \ \imp\ 
   (\satf \for{\Veb{\th}^s} \land \for{\Veb{\th}^i}).$$
Since the set of scalar formulae is fixed (does not vary with $b$), the
above depends only on \th. We therefore render it as a wff as follows:
\newsavebox{\indSuff}
\sbox{\indSuff}{Th$(\th)$}
\bleqn{\usebox{\indSuff}}
\parbox{2.8in}{
 $\fa b > 0: (\ex \bar{z}, a: \for{\Veb{b}^s} \land \for{\Veb{\th}^i}) \imp
                    (\ex \bar{z}, a:  \for{\Veb{\th}^s} \land \for{\Veb{\th}^i}).$
}
\eleqn
This is still not quite a wff, since it is not closed: it depends on \th. We
cannot add a $\ex \th$ quantifier at the beginning, since the form of the
formula changes with \th (same problem we had above with $b$).
So, we check Th(\th) for values of \th starting from 1 and
incrementing. Hence, we find the smallest value of \th which works, as desired.

Clause~\ref{project-val-user} must be assumed as an axiom, since it is a
restriction on user behavior:

\textbf{User-Consistency}:
Let $b > \th$, and let $\Veb{\th} = \Veb{b} \pj \ethb{\th}$. Then 
the user assigns the same classification (\vtt or \vff) to 
all $\asg_b \in \md{\Veb{b}}$, and $\asg_{\th} \in \md{\Veb{b} \pj \ethb{\th}}$.


From Clauses~\ref{project-val-satis} and~\ref{project-val-user},
we obtain:
\begin{quote}
  for all $b \ge \th$, if some $\asg \in \md{\Veb{b}}$ exists, then some
  $\asg_{\th} \in \md{\Veb{\th}}$ exists, and the user gives the same
  answers to the queries $\asg \in \vtt$ and $\asg_{\th} \in \vtt$.
\end{quote}
Hence we can present $\asg_{\th} \in \vtt?$ to the developer rather than $\asg \in \vtt?$

\paragraph{Example: array search.}
Let \eth be the equivalence theory for array search given above. Then 
for \ethb{b}, 
the scalar formulae are
$\l = r$, $\l < r$, $\l < r-1$,
$\l \ge 0$, $\l \le b-1$,
$r \ge 0$, $r \le b-1$, 
and the indexed formulae are 
$\l=0, \l=1, \ldots \l=b-1$,
$r=0, r=1, \ldots, r=b-1$, 
$a[0]=e, a[1]=e, \ldots, a[b-1]=e$.

Now for \ethb{\th} with $\th < b$, 
the scalar formulae are
$\l = r$, $\l < r$, $\l < r-1$,
$\l \ge 0$, $\l \le \th-1$,
$r \ge 0$, $r \le \th-1$, 
and the indexed formulae are 
$\l=0, \l=1, \ldots \l=\th-1$,
$r=0, r=1, \ldots, r=\th-1$, 
$a[0]=e, a[1]=e, \ldots, a[\th-1]=e$.

Note that the indexed formulae of \ethb{\th} are a subset of those of \ethb{b},
while the scalar formula are not: they result by substituting \th for $b$.

Let \Veb{b} be an assignment to 
$\l = r$, $\l < r$, 
$\l \ge 0$, $\l \le b-1$,
$r \ge 0$, $r \le b-1$, 
$\l=0, \l=1, \ldots \l=\th-1$,
$r=0, r=1, \ldots, r=\th-1$, 
$a[0]=e, a[1]=e, \ldots, a[\th-1]=e$.

Let \Veb{\th} be an assignment to 
$\l = r$, $\l < r$, 
$\l \ge 0$, $\l \le \th-1$,
$r \ge 0$, $r \le \th-1$, 
$\l=0, \l=1, \ldots \l=\th-1$,
$r=0, r=1, \ldots, r=\th-1$, 
$a[0]=e, a[1]=e, \ldots, a[\th-1]=e$.

Th(\th) states that if  \Veb{b} is satisfiable, then so is 
\Veb{\th}. Suppose that \Veb{b} assigns true to 
$\l < r$, 
$\l \ge 0$, 
$r \le b-1$,
and truth values to other formulae so that \Veb{b} is satisfiable.
Then \Veb{\th} assigns true to 
$\l < r$, 
$\l \ge 0$, 
$r \le \th-1$,
and must also be satisfiable. This requires $\th \ge 2$.
If we had included $\l < r-1$ in $\eth$, \eg to require at least one array
element between the left and right boundaries, then we would have $\th \ge 3$.
We validated this by composing Th$(\th)$ manually and submitting to
Z3, with values 1,2,3 for $\th$. Th$(\th)$ was not valid for
1,2, and was valid for 3, as expected.

These lower bounds on \th show that we need \ethb{\th}
to have array sizes sufficiently large to be able to represent all satisfiable
assignments to the formulae in any \ethb{b}.

\section{Vocabulary and Quantifier Construction}
\label{sec:quantconstruct}
We present our vocabulary construction method, and then
present three complementary methods to construct 
quantified formulae. 
Our vocabulary construction method takes as input:
\be
\i A type theory \tth expressed as a set of,
variables $X$ and a map from $X$ to scalar and array
types,
\i A set of literal constants $L$ such as $0,1,\mathit{true},$ and $\mathit{false}$,
\i A Presburger and index operations alphabet $\Sigma=\set{
\mathit{index},\mathit{bound},=,<=,+,-,*,[]}$,
\i A grammar $G\subseteq X\times X\cup L \times 2^{\Sigma}$, denoting the allowed operations between variables, 
\i A bound $K$ expressing the maximum number of allowed operations in a clause
\ee

The method then traverses the grammar $G$ and builds
\voc to be the set of 
all Boolean formulae with up to $K$ operators. 
We pass \voc to \constFor.
This relieves the user from providing both \voc and 
\eth since we can also use an extended grammar $G'$ 
and a larger bound $K'>K$ for \eth. 

\subsection{Quantified formula construction}
\label{sec:quantformconst}

The work in~\cite{Furia2010} and~\cite{Brdly06} discuss 
decidable fragments of the theories of sequences and arrays. 
The {\em array property theory}
$\exists\forall_i\tau$ presented in~\cite{Brdly06} allows
restricted existential and universal quantification of 
the form 
$\forall
\bar{x}. \phi(\bar{x}) \rightarrow \psi(\bar{x})$.
$\exists\forall_i\tau$ limits universal quantification to the
variables used in index terms, limits Presburger arithmetic
expressions used in $\phi$ for quantified variables, and 
allows Presburger arithmetic and Boolean operations 
in $\phi$ and $\psi$.
$\exists\forall_i\tau$ is defined by a grammar which restricts the
syntax of formulae appropriately.

Satisfiability of $\exists\forall_i\tau$ is polynomially 
reducible to satisfiability of quantifier free, 
uninterpreted functions, equality
theory (QF-EUF) with additional free variables each of 
which replaces one universally quantified 
variable~\cite{Brdly06}.

The quantified formula construction method takes an 
additional bound $N$ from the user denoting
the maximum number of quantified variables.
The method constructs the set $X'=X\cup X_N$ where $X_N$ 
has up to $N$ fresh scalar variables,
and constructs the set $G'$ by adding rules to $G$
that relate the fresh variables in $X_N$ to
the array variables in $X$. 
The method then traverses the grammar $G'$ and builds
$\exists\forall_i^N\tau^K$ the set of
all Boolean formulae with up to $K$ operators. 
The construction of the 
$\exists\forall_i^N\tau^K$ theory is further detailed 
online$^{\ref{fn:online}}$.

If the grammar provided by the user is within the grammar
of~\cite{Brdly06} and~\cite{Furia2010}, then 
the theory $\exists\forall_i^N\tau^K$ 
is a subset of $\exists\forall_i\tau$;
it is reducible to QF-EUF
which renders queries to the SMT solver decidable.
$\exists\forall_i^N\tau^K$ is also powerful enough 
to express
formulae under the array property and the list property
theories with up to $N$ quantifiers and $K$-operation 
Boolean terms. 
We leave the grammar restriction as a user choice
to benefit from other decidable theories covered by the 
SMT solvers. 

We use $\exists\forall_i^N\tau^K$ as
our vocabulary and we run \constFor\ to construct the 
desired formula. 
When querying the user, we hide the values of the 
$X_N$ variables. 
In practice, if a presented assignment is not enough
to judge \vtt or \vff, 
this is an indication that the generated
vocab is not adequate and that the user should 
increase either $N$ or $K$. 

Once \constFor\ returns \Fo, 
we deskolmnize it and construct
$\exists \bar{X_N}.\Fo_{k-1}$ by 
existentially quantifying the $X_N$ variables 
in $\Fo$ that were 
not part of the original type theory
provided by the user. 
This works well, since \Fo is a disjunction of vocabulary evaluations
(each of which is a conjunction of clauses)
and existential quantification distributes through disjunction.
This allows us to move the $\exists$ to the beginning of $\Fo$, as in 
$\exists \bar{X_N}.\Fo_{k-1}$.

For example, 
the method took the \cci{Input} type theory and grammar 
in Figure~\ref{f:einals} 
that specified an array $a$, two bounds $\l$ and $r$, and
a scalar $e$ denoted of element sort by the \cci{(a,e,=)} grammar 
rule, extended the variable set with $i$, added 
the rule \cci{(a,i,index)} to the grammar, generated a 
vocab with $K=1$ and called \constFor\ to construct the 
formula \cci{eina(a,left,right,e)} specifying that 
$e$ is in $a$ between $\l$ and $r$ inclusive.

The method also took the type theory that specified
$a$ as an array, and one grammar rule \cci{(a,a,<=)}
rule that allowed elements of $a$ to be compared 
with each other, injected the variable $i$ as an index
to $a$, constructed a vocab that with $k=3$ that included 
a Presburger index term $i+1$,
called \constFor\ to generate the formula 
\cci{notsorted(a)}. 
The formula \cci{notsorted(a)} can be negated to express
\cci{sorted(a)}.
The generated formulae can then be used as
vocabulary clauses
in the construction of other formulae.

Note that, the same method can be applied to obtain 
universally
quantified formulae with a variant of \constFor\ where
we construct $\neg\Fo$, the complement of \Fo,
where we start with \true\ instead of \false,
and proceed to trim the formula $\neg\Fo$ by conjunctions 
of formulae corresponding to vocab evaluations 
deemed \vff by the user; instead of adding disjunctions
of those deemed \vtt to \Fo.
Therefore, a \cci{sorted(a)} can be generated directly 
as a universally quantified formula. 

\subsection{Hierarchical and Incremental Construction}

Our method builds the formula in incremental steps from 
bottom to top. 
Let $\Fo_{k-1}$ be a formula generated at level $k-1$.
We deskolemize it and generate
$\exists \bar{x}.\Fo_{k-1}$
where $\bar{x}$ represents the variables 
in $\Fo_{k-1}$ that
are not part of $\tth_k$, the type theory of $\Fo_k$.
We then introduce $\exists \bar{x}.\Fo_{k-1}$
as a clause $C$ in the vocabulary of $\Fo_k$. 

Consider the formula \cci{eina(a,left,right,e)}
from Figure~\ref{f:einals}
constructed to express the existence of 
element $e$ in array $a$ between bounds $\l$ (\cci{left}) and $r$ (\cci{right}) 
in the process of constructing \Fo; a postcondition
for linear search.
We deskolemnize and introduce 
\cci{eina(a,left,right,e)}$=\exists i.\Fo_{k-1}$
as a clause $C$ in the construction of \Fo. 
We use $C$ along with other vocabulary
clauses, resulting in 
$(0\le \l \le r \le |a|-1) \land \big(
(rv\not= -1 \land e=a[rv] \land eina(a,\l,r,e))
\lor
(rv=-1 \land \neg eina(a,\l,r,e)) \big)$
where the negation introduces a universal quantifier. 
Figure~\ref{f:einals} shows a sample output for
the linear search postcondition using the incremental 
method.

\begin{figure}[t!]
\center{
\begin{Verbatim}[fontsize=\relsize{-2},frame=topline,framesep=4mm,label=\fbox{eina(a,left,right,e)}]
Input: theory eina {
  int [] a; int left; int right; int e;
  grammar { (a,left,bound); (a,right,bound); (a,e,=);} }
... 
Spec: exists i. 0<= left and left <= right 
        and right <= a.size - 1 
        and left <= i and i <= right and a[i]=e

\end{Verbatim}

\begin{Verbatim}[fontsize=\relsize{-2},frame=topline,framesep=4mm,label=\fbox{not-sorted(a)}]
Input: theory not-sorted {
  int [] a;
  grammar { (a,a,<=);}}
... 
Spec: exists i. 0 <= i and i <= a.size - 1 and 
            0 <= i + 1 and i + 1 <= a.size - 1  and
            not (a[i] <= a[i+1])

\end{Verbatim}

\begin{Verbatim}[fontsize=\relsize{-2},frame=topline,framesep=4mm,label=\fbox{array-search-post(a,left,right,e,rv)}]
input: theory ls-post {
  int [] a; int left; int right; int e; int rv; 
    bool eina(a,left,right,e);
  grammar { (a,left,bound); (a,right,bound); 
    (a,rv,index); (a,e,=);} }
...
Spec: 0 <= left and left <= right and right <= a.size - 1 
      and (rv = -1  and !eina(a,left,right,e) or
           rv != -1 and eina(a,left,right,e) and a[rv]=e)
\end{Verbatim}
}
\caption{Sample formulae generated using \constFor}
\label{f:einals}
\end{figure}

We could have written the linear search 
postcondition directly without using the \cci{eina}
clause. However, that results in 
more queries to the user and the SMT solver. 
Building formulae incrementally also builds a rich library
of accurate reusable specifications. 

%

\section{Implementation}
\label{sec:implementation}
The implementation of \constFor\ and \mkAdeq\ is available 
online~\footnote{\label{fn:online}\url{
http://webfea.fea.aub.edu.lb/fadi/dkwk/doku.php?id=speccheck}}. 
The tool \cci{sc} implements \constFor\ and
takes a type theory $\tth$ as a set of variable
declarations.
It also takes a vocabulary $\voc$ as a set of SMT formulae.
Optionally, it takes a grammar that relates the variables to each
other, and generates a vocabulary $\voc_g$ from the type theory and the
grammar.
The user has the option to produce a quantifier free equivalent of the
\voc or $\voc_g$ if they are under the array or the list
property theories~\cite{Brdly06,Furia2010} 
to guarantee successful SMT calls.

The user also specifies other options such as the maximum number 
of quantifiers, the type of the quantifier,
and the maximum number of operations per generated vocabulary clause.
Upon successful termination, the tool uses 
ESPRESSO~\cite{BraytonEspresso1984} and 
ABC~\cite{ABCMeshinkoDATE2013}, 
logic synthesis tools, to simplify the specification.
The simplified specification is then presented to the user. 

The tool \cci{ma} takes a type theory $\tth$, 
an equivalence theory $\eth$, and
a vocabulary $\voc$ and augments $\voc$ if needed so 
that it is adequate as described in \mkAdeq. 
All tools use the C++ api of the Z3 SMT solver~\cite{Z309}.

\section{Results}
\label{s:examples}

\begin{table*}[bt]
\centering
\caption{Results of user experiments with \cci{sc} }
\resizebox{1.4\columnwidth}{!}{
\begin{tabular}{p{1.5cm}|c|c|c|c|c|c|c|c|c}
  Spec & User & Attempts & Clauses & Accuracy & User & SMT & Used & SMT time & Total \\ 
       &      &           &          &          & Queries & Queries & PA & $10^{-6}$(seconds) & time (s)\\ \hline \hline
\multirow{7}{1.5cm}{eina, user vocab: 3 clauses, inject quantifier}
 & \multirow{5}{*}{student}  
    & 3 & 5 & correct & 8 & 14 & 0 & 4190 & 821 \\ 
 &  & 1 & 3 & missed range on i & 8 & 8 & 2 & 3299 & 715 \\ 
 &  & 1 & 3 & missed range on i & 5 & 5 & 3 & 2302 & 748.09 \\ 
 &  & 2 & 5 & correct & 8 & 10 & 1 & 2991 & 542.213 \\ 
 &  & 1 & 3 & correct & 8 & 8 & 0 & 2667 & 76 \\ 
\cline{2-10}
 & \multirow{2}{*}{expert}
    & 1 & 5 & correct & 8 & 10 & 1 & 2785 & 543 \\ 
    &  & 1 & 5 & correct & 8 & 10 & 1 & 2855 & 588 \\ 
\hline
\multirow{6}{1.5cm}{ eina , type theory ,  generated}
& \multirow{4}{*}{student} 
    & 2 & 20 & correct & 125 & 182& 7 & 57558 & 2329 \\ 
 &  & 3 & 20 & correct & 243 &323 & 4&113870  & 1934.72 \\ 
 &  & 1 & 20 & correct & 84 & 153 & 7 & 43232 & 1468.87 \\ 
 &  & 2 & 20 & correct & 59 & 117 & 9 & 30699 & 1584.01 \\ 
 &  & 2 & 20 & correct & 31 & 69 & 11 & 12194 & 746.3 \\ 
\cline{2-10}
 & \multirow{2}{*}{expert}
    & 1 & 20 & correct & 44 & 88 & 9 & 20785 & 1096 \\ 
 &  & 1 & 20 & correct & 33 & 71 & 10 & 13042 & 770 \\ 
\hline
\multirow{6}{1.5cm}{ linear search, type theory, incremental} 
& \multirow{7}{*}{student}
    & 2 & 20 & correct & 30 & 79 & 6 & 19296 & 717 \\ 
 &  & 3 & 20 & correct & 243 & 323 & 4 & 113870 & 1934.72 \\ 
 &  & 1 & 5 & missed-rv=-1 & 25 & 29 & 1 & 10355 & 230.488 \\ 
 &  & 1 & 5 & correct & 10 & 11 & 3 & 3500 & 496.174 \\ 
 &  & 1 & 5 & correct & 32 & 32 & 0 & 3138 & 336.851 \\ 
 &  & 1 & 5 & missed-rv=-1 & 9 & 10 & 3 & 3970 & 256.58 \\ 
 &  & 2 & 20 & correct & 37 & 91 & 5 & 15321 & 362.972 \\ 
\cline{2-10}
 & expert 
    & 1 & 20 & correct & 28 & 72 & 7 & 12241 & 543 \\ 
\hline
\multirow{6}{1.5cm}{sorted, type theory}  
 & \multirow{5}{*}{student}
    & 1 & 7 & correct & 19 & 23 & 2 & 7529 & 406 \\ 
 &  & 1 & 7 & missed-i+1 bound & 36 & 40 & 1 & 13032 & 234.696 \\ 
 &  & 2 & 7 & correct & 26 & 30 & 2 & 10178 & 221.476 \\ 
 &  & 2 & 7 & correct & 36 & 40 & 0 & 13232 & 181.09 \\ 
 &  & 2 & 7 & correct & 11 & 15 & 5 & 3506 & 270 \\ 
\cline{2-10}
 & expert & 1 & 7 & correct & 12 & 12 & 4 & 5598 & 420 \\ 
\hline
\end{tabular}
}
\label{t:exp}
\end{table*}

We conducted a user experiment using the array search 
and the array sorted examples. 
Table~\ref{t:exp} shows the results. 
Eight volunteer 
students and two logic design experts were asked 
to construct specifications 
using \cci{sc}. 
They were trained to use \cci{sc} with simple examples
that specify orders between scalars. 
Then they were given \cci{sc} with an assignment sheet 
that instructed to specify the following. 
\be
\i \cci{eina(a,left,right,e)} using a startup vocab.
\i \cci{eina(a,left,right,e)} using a type theory and a grammar. 
\i Incrementally specifying the array search property. 
\i sorted(a).
\ee

The tool currently has no undo facility, so 
users aborted the run when they provided 
an unintentional (mistaken) answer. 
They were not allowed to attempt again once they achieved
a constructed specification.
The attempts column reflects the number of aborted
attempts by the user, plus the final attempt. 

We explained the partial assignment optimizations to 
the users, and we warned them against using them.
They still used them to save time, 
especially with the generated vocabulary.
Users who 
specified partial assignments 
made mistakes more often and 
forced the tool to ignore valid \vtt assignments.
Some users, who used the optimizations to save time, 
ended up calling the SMT solver less often, but spent more
total time discerning their optimizations. 
All users, including the two experts,
reported their surprise on how much easier it was to construct the 
\cci{sorted} property compared to the \cci{search} property using the tool. 
The accuracy of the constructed formulae was jointly assessed by the
users and the authors.
Some users left early, due to timing constraints.

The authors of the tool also validated the tool 
by writing accurate specifications for memory allocation and deallocation,
linked list validity, binary search tree properties, red black binary search
tree properties, rooted index tree properties, and a text justify example. 
The assignment sheet and samples from the logged results are available
online$^{\ref{fn:online}}$.


\remove{
We present in the appendices two examples: array search and text justification.
For array search, we obtain precondition 
$0 \le l \land l \le r \land r \le |a|$, where $\l, r$ are the left,
rigth boundaries, respectively of the section of $a$ to be searched. 
We obtain postcondition 
$(rv = -1 \land (\fa i : \l \le i \le r : a[i] \ne e) \lor (0 \le l \le rv \le r < |a| \land a[rv] = e)$,
where $e$ is the value being searched for, and $rv$ is the return
value: -1 if $e$ is not present in $a$ between positions $\l$ and $r$ inclusive, and otherwise a location of $e$ between
$\l$ and $r$.
We also show how an inadequate vocabulary \set{0 \le l, \l \le r} is corrected by \m
}

\section{Related work }
\label{s:related}



\label{sec:related}

The methods in \cite{HWT03,HST98,HKLAB98,KKK95,CABetal98} work 
by writing the specification and
then attempting to verify if it is
accurate using animation, execution, model-checking, etc.
We go in the other direction: we write the
specification from the behaviors, so that the specification is
accurate by construction. 
\remove{In \cite{PP03}, a method for writing
trace-based specifications is presented. The behaviors are sequences
of atomic events. The technique is suitable for specifying reactive modules as
``black boxes'' that interact with an environment via events. Our
method, in contrast, views programs as white boxes, and our
specification are pre/postcondition pairs over the program state. Our
method is intended for transformational programs that perform nontrivial
computations on data, rather than reactive modules where control is
the major issue.}
A method of writing temporal-logic based specifications using event
traces (``scenarios'') is presented in \cite{LW98}. It applies
to reactive systems and stresses control rather than data.
In \cite{F89}, a method for refining an initially simple specification
using informal ``elaborations'' is presented. 
A method of checking software cost reduction (SCR) specifications for consistency 
is presented in \cite{HJL96}. 
Zeller in~\cite{zeller-spin-2012} discusses writing specifications as models
discovered from existing software artifacts of relevance to the desired
functionality. 
In none of the above is there an analogue to our construction of
preconditions and postconditions as formulae of first order logic.

In \cite{jha-icse10}, an oracle-based method computes a loop free
program that requires a distinguishing constraint and an I/O behavior
constraint. The method either synthesizes a program or claims the
provided components are insufficient.  We differ in that we build
specifications in first order logic with quantifiers, we do not
require the ``correctness'' of $\eth$, and we correct $\voc$ when it is
not adequate.

The SPECIFIER \cite{MH91} tool constructs formal specifications of
data types and programs from informal descriptions, but uses schemas,
analogy, and difference-based reasoning, rather than 
input-output behaviors.  Larch \cite{GGH90}
enables the verification of claims about specifications,
which improves the confidence in the specification's accuracy.
In \cite{MS03}, a method for testing preconditions, postconditions,
and state invariants, using mutation analysis, is presented. 

\section{Conclusion}
\label{s:conclusion}
We presented a method to construct a formal specification, 
given (1) an adequate formal vocabulary, 
and (2) interaction with a user who can accurately 
classify behaviors.  
We illustrated our method with examples
and evaluated it by conducting user experiments.  
We illustrated our method by constructing specifications 
for array search, binary search, red black binary search
trees, and root indexed trees. 
We also conducted a controlled experiment with 
senior undergraduate and graduate students and logic design
experts.

\clearpage
\bibliographystyle{plain}
\bibliography{specConstruct}

\end{document}